\documentclass[preprint,12pt]{elsarticle}

\usepackage{amssymb}

\journal{Applied Mathematical Modelling}

\begin{document}

\begin{frontmatter}




\title{Three-dimensional sand ripples as the product of vortex instability \tnoteref{label_note_copyright} \tnoteref{label_note_doi}}

\tnotetext[label_note_copyright]{\copyright 2016. This manuscript version is made available under the CC-BY-NC-ND 4.0 license http://creativecommons.org/licenses/by-nc-nd/4.0/}

\tnotetext[label_note_doi]{Accepted Manuscript for Applied Mathematical Modelling, v. 37, p. 3193-3199, 2013, doi:10.1016/j.apm.2012.07.027}


\author{Erick de Moraes Franklin}

\address{Instituto de Engenharia Mec\^anica - Universidade Federal de Itajub\'a\\
e-mail: erick@unifei.edu.br\\
Av. BPS, 1303 - Itajub\'a - MG - CEP: 37500-903\\
Brazil}

\begin{abstract}
Three-dimensional sand ripples can be observed under steady liquid flows in both nature and industry. Some examples are the ripples observed on the bed of rivers and in petroleum pipelines conveying sand. Although of importance, the formation of these patterns is not completely understood. There are theoretical and experimental evidence that aquatic ripples grow from two-dimensional bed instabilities, so that a straight vortex is formed just downstream of their crests. The proposition of Raudkivi (2006), that three-dimensionality has its origin in a vortex instability, is employed here. This paper presents a linear stability analysis of the downstream vortex in order to obtain the transverse scales of three-dimensional ripples. The obtained wavelength is compared with experimentally observed ripples.
\end{abstract}


\end{frontmatter}



\section{INTRODUCTION}

When a liquid flows over a bed of grains, some of them can be entrained by the flow. If the shear stresses caused by the fluid are within some limits, the grains are transported as a mobile granular layer known as bed-load \cite{Bagnold_1}, which stays in contact with the fixed part of the bed. Under some conditions, a granular bed may be unstable, generating two-dimensional bedforms that evolve into different kinds of ripples and dunes \cite{Engelund_Fredsoe}. Two-dimensional ripples and dunes have their crests aligned in the direction transverse to the mean flow, so that their shape is characterized by a longitudinal wavelength.

In the last decades, much work was devoted to understanding the formation of two-dimensional ripples and dunes under steady flows \cite{Kennedy, Reynolds, Engelund_1, Fredsoe_1, Richards, Elbelrhiti}. In these studies, the mechanisms of the instability were explained, and linear stability analyses were conducted. However, only some analyses were concerned with saturation and three dimensionality.

Three-dimensional sand ripples are frequently observed under steady liquid flows in both nature and industry. Some examples are the ripples observed on the bed of rivers and open-channels, but also in petroleum pipelines conveying sand. Although of importance, the origin of these three-dimensional patterns is not completely understood.

Langlois and Valance (2005) \cite{Langlois_2} performed a three-dimensional stability analysis of a granular bed sheared by a laminar flow, with infinite depth and long-wavelength approximations. In the linear phase, they found that the most unstable mode is longitudinal, corresponding to two-dimensional ripples. They also found the existence of oblique modes, that are coupled in the nonlinear regime to the longitudinal mode and give rise to three-dimensional ripples. However, Langlois and Valance (2005) \cite{Langlois_2} did not take into account the effects of the recirculation vortex, that appears downstream of each ripple in the nonlinear phase.

In a recent paper, Franklin (2010) \cite{Franklin_4} presented a linear stability analysis of a granular bed sheared by a turbulent liquid flow, without free surface effects. The most unstable wavelength, growth rate and celerity of the two-dimensional forms were obtained, and compared well with experimental results. Giving the linear character of the analysis and the absence of free surface effects, the obtained solutions are valid only for the initial phase of the instability.

\begin{sloppypar}
In order to understand the evolution of two-dimensional bedforms, Franklin (2011) \cite{Franklin_5} and Franklin (2012) \cite{Franklin_6} presented nonlinear stability analyses with the same scope as \cite{Franklin_4}. The employed approach was the weakly nonlinear analysis \cite{Landau_Lifshitz, Schmid_Henningson, Drazin_Reid, Livre_Charru} and it was showed that the instabilities saturate after the initial linear phase, attenuating their growth rate and maintaining the same wavelength. Also, following the proposition of Fourri\`ere et al. (2010) \cite{Fourriere_1} that dunes are secondary instabilities formed from the coalescence of ripples, Franklin (2012) \cite{Franklin_6} included free surface effects and obtained the length-scale and the celerity of dunes.
\end{sloppypar}

There is experimental evidence that the three-dimensional patterns are formed after the linear phase, when the wavelength of bedforms is saturated \cite{Raudkivi_2, Franklin_3}. In this picture, the explanation given by Raudkivi (2006) \cite{Raudkivi_2} is the most plausible: the formation of three-dimensional bedforms is directly related to the structures of the fluid flow. In particular, he proposed that deformations of the downstream vortex could lead to three-dimensional ripples. This vortex, initially parallel to the crest line, is generated only when the bedforms are relatively high. To the author's knowledge, a mathematical model taking into account this mechanism has not yet been proposed for the ripples.

This paper explores the idea proposed by \cite{Raudkivi_2}. Initially considering two-dimensional ripples and their recirculation vortices, a linear stability analysis is made of a vortex. The longitudinal wavelength of the vortex instability is then assumed to scale with the transverse wavelength of three-dimensional ripples. The liquid flow is assumed to generate strong vortices, so that it is considered as turbulent. In order to conduct the stability analysis, the vortex is modeled as a line vortex in the transverse direction. The most unstable mode is identified and the obtained wavelength is compared to experimental data.

The next section presents a linear stability analysis of the line vortex downstream of a two-dimensional ripple. The following section discusses the results and compares them with some experimental data. The conclusion section follows.

\section{STABILITY ANALYSIS}
\label{section:analysis}

A linear stability analysis of the vortex downstream of a two-dimensional ripple is presented here. This vortex is aligned in the direction transverse to the mean flow, and it is modeled as a line vortex. The deformations of the line vortex are assumed to scale with transverse structures of the bedforms.

The initial picture is a longitudinal turbulent flow over a two-dimensional ripple, and a line vortex downstream of its crest. This line vortex shall then extend to the infinity, or end at the walls in the case of finite channels \cite{Batchelor}. Deformations of this line vortex have two contributions: the velocity induced by the line vortex on itself, called auto-induction; and the external shear flow.

\begin{figure}
  \begin{center}
    \includegraphics[width=0.6\columnwidth]{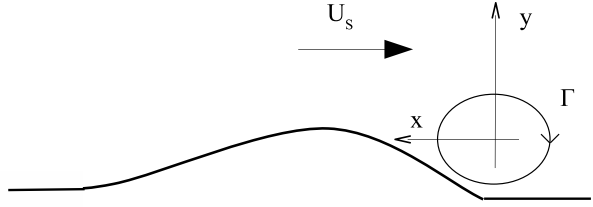}
    \caption{Layout of the vortex downstream a ripple. $U_S$ is the velocity of the flow above the vortex, $x$ is the longitudinal direction, $y$ is the vertical direction and $\Gamma$ is the intensity of the line vortex.}
    \label{fig:scheme1}
  \end{center}
\end{figure} 

The Biot-Savart law of induction determines the velocity induced by a vortical structure on a given position $\vec{r}$. For a line vortex of intensity $\Gamma$, the induced velocity field is \cite{Batchelor, Saffman}

\begin{equation}
\vec{u}(\vec{r})\,=\,-\frac{\Gamma}{4\pi}\int{\frac{\vec{D}\times\vec{d\ell}(\vec{r}_\ell)}{| \vec{D}|^3}}
\label{biot_savart}
\end{equation}

\noindent where $\vec{r}_\ell$ are the points of the line vortex, $\vec{D}\,=\,\vec{r}-\vec{r}_\ell$, $\vec{d\ell}$ is a segment of the line vortex and $\vec{u}(\vec{r})$ is the induced velocity at $\vec{r}$. In the case of a straight line vortex, auto-induction is absent because all the points $\vec{r}$ lie outside the line vortex. However, in the presence of curvature some of the $\vec{r}$ points lie at the line vortex, so that it induces a velocity on itself.

Let $\zeta$ be a curve in space defined by the function $\vec{L}(s,t)$, where $s$ is a segment of the curve and $t$ is the time. Consider the Frenet-Serret frame, given by the tangent $\vec{\tau}$, the normal $\vec{n}$ and the binormal $\vec{b}$ unit vectors with respect to $\vec{L}(s,t)$. An approximation for the auto-induced velocity $\dot{\vec{L}}_{AI}(s,t)$ can be obtained by applying Eq. \ref{biot_savart} to a bent line-vortex of radius $R$, as suggested by \cite{Saffman, Guyon}

\begin{equation}
\dot{\vec{L}}_{AI}(s,t)\,\sim\,\frac{\beta}{R}\vec{b}
\label{auto-induction}
\end{equation}

\noindent where

\begin{equation}
\beta\,=\,-\frac{\Gamma}{4\pi}\log\left( \frac{\xi}{R} \right)
\label{beta}
\end{equation}

\noindent and $\xi$ is the radius of the vortex core, $\xi\ll R$.

\begin{figure}
  \begin{center}
    \includegraphics[width=0.6\columnwidth]{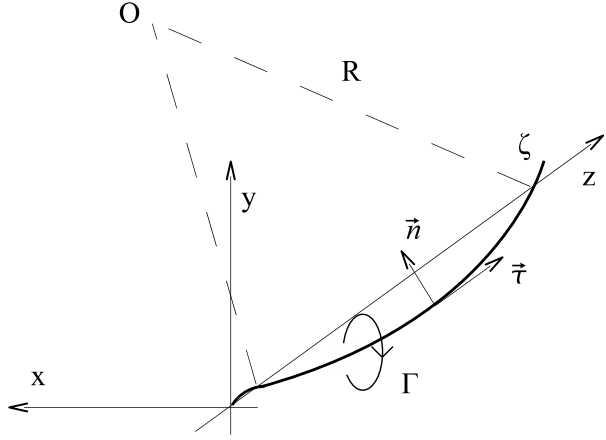}
    \caption{Line vortex in the transverse direction under small perturbations. $R$ is the radius of curvature of the line vortex, $O$ is the origin of the curvature, and $\vec{n}$ and $\vec{\tau}$ are the normal and the tangent unit vectors, respectively.}
    \label{fig:scheme2}
  \end{center}
\end{figure} 

The mean velocity of the shear flow $\vec{U}_S(\vec{r})$ is approximated as varying linearly in the vertical direction $y$, which is a reasonable assumption in the near vortex region:

\begin{equation}
\vec{U}_S(\vec{r})\,\approx\,-\alpha y \vec{i}\,+\,0\vec{j}\,+\,0\vec{k}
\label{Us}
\end{equation}

\noindent where $\vec{i}$, $\vec{j}$, and $\vec{k}$ are the unit vectors of the Cartesian coordinate system in the ripple frame of reference. The shear velocity acts directly on the line vortex, so that its contribution to the vortex deformation is

\begin{equation}
\dot{\vec{L}}_{S}(s,t)\,\approx\,-\alpha y \vec{i}
\label{shear}
\end{equation}

Finally, the local displacement of the line vortex, that gives its deformation, is

\begin{equation}
\dot{\vec{L}}(s,t)\,=\,\dot{\vec{L}}_{AI}(s,t)\,+\,\dot{\vec{L}}_{S}(s,t)
\label{deformation}
\end{equation}

A linear stability analysis of the line vortex, based on Eqs. \ref{auto-induction} to \ref{deformation}, is presented next. The deformation of the line vortex is divided in a basic state, whose variables are $O(1)$, and in vortex perturbations, which are $O(\epsilon)$, $\epsilon \ll 1$. The equations are then developed and higher order terms are neglected.

\subsection{Basic state}

The basic state is a straight line vortex, aligned in the transverse direction. In this case, all the segments $s$ are in the transverse direction, $\vec{L}(s,t)\,=\,s\vec{k}$, and auto-induction is absent.

\subsection{Perturbations of the line vortex}

Perturbations of the line vortex are considered as small deformations in the longitudinal and vertical directions

\begin{equation}
\vec{L}(s,t)\,=\,x(s,t)\vec{i}\,+\,y(s,t)\vec{j}\,+\,s\vec{k}
\label{perturbation}
\end{equation}

\noindent where $x(s,t)/s$ and $y(s,t)/s$ are $O(\epsilon)$. A system of equations for the evolution of perturbations can be obtained by differentiating Eq. \ref{perturbation} with respect to time, and comparing it with Eq. \ref{deformation}. However, one needs to consider only deformations in directions perpendicular to the line vortex:

\begin{equation}
\dot{\vec{L}}(s,t)\times\vec{\tau}\,=\,\frac{\beta}{R}\vec{b}\times\vec{\tau}\,+\,\vec{U}_S(\vec{r})\times \vec{\tau} 
\label{perturbation_equation}
\end{equation}

Developing Eq. \ref{perturbation_equation} and neglecting higher order terms, the following system is obtained

\begin{equation}
\begin{array}{c} \dot{x}\,=\,-\beta y''-\alpha y \\ \, \\ \dot{y}\,=\,\beta x'' \\ \end{array}
\label{system}
\end{equation}

The solutions to this linear system consist of plane waves. These solutions can be found by considering the normal modes

\begin{equation}
\begin{array}{c} x\,=\,Ae^{i\left( ks-\Omega t\right) }\,+ c.c. \\ \, \\ y\,=\,Be^{i\left( ks-\Omega t\right) }\,+ c.c. \\ \end{array}
\label{normal_modes}
\end{equation}

\noindent where $k\in \mathbb{R}$, $k=2\pi\lambda^{-1}$ is the wavenumber in the $s$ direction, $\lambda \in \mathbb{R}$ is the wavelength in the $s$ direction, $A\in \mathbb{C}$ and $B\in \mathbb{C}$ are the amplitudes and $c.c.$ stands for complex conjugate. Let $\Omega\in \mathbb{C}$, $\Omega\,=\,\Omega_r\,+\,i\Omega_i$, where $\Omega_r\in \mathbb{R}$ is the angular frequency and $\Omega_i\in \mathbb{R}$ is the growth rate. By inserting the normal modes in Eq. \ref{system}, one obtains

\begin{equation}
\begin{array}{c} \left( i\Omega_r-\Omega_i\right) A\,+\,\left( \beta k^2-\alpha\right)B\,=\,0 \\ \, \\ \beta k^2A\,-\,\left( i\Omega_r-\Omega_i\right) B\,=\,0 \\ \end{array}
\label{system2}
\end{equation}

The existence of non-trivial solutions to this system requires its determinant to be zero. This gives

\begin{equation}
\begin{array}{c} \Omega_r^2-\Omega_i^2\,=\,\beta k^2\left( \beta k^2-\alpha\right)\,=\,0 \\ \, \\ 2\Omega_r \Omega_i=\,0 \\ \end{array}
\label{system3}
\end{equation}

A long-wavelength approach is adopted here (this will be justified \textit{a posteriori} in Section \ref{section:comparison}), so that $\beta k^2-\alpha\,<\,0$ and the only possible solutions are

\begin{equation}
\Omega_r\,=\,0
\label{frequency}
\end{equation}

\begin{equation}
\Omega_i\,=\,\pm\sqrt{\beta k^2\left(\alpha -\beta k^2\right) }
\label{growth_rate}
\end{equation}

The solutions are waves with zero celerity and with negative (stable branch) and positive (unstable branch) growth rates. The growth of three-dimensional ripples corresponds to $\Omega_i>0$. 

\subsection{Analysis}

The linear analysis implies that, in a time scale $O(\Omega_i^{-1})$, the unstable modes grow exponentially. A maximum in the curve $\Omega_i(k)$ corresponds then to the most unstable mode, i.e., the one that grows faster than all the others. In this mode, $\partial\Omega_i/\partial k=0$ and $\partial^2\Omega_i/\partial k^2|_{k=k_{max}}<0$, where $k_{max}$ is the most unstable wavenumber. When $\Omega_i>0$, the latter is verified and the former gives

\begin{equation}
k_{max}\,=\,\sqrt{\frac{\alpha}{2\beta}}
\label{k_max}
\end{equation}

\noindent and

\begin{equation}
\lambda_{max}\,=\,2\pi\sqrt{\frac{2\beta}{\alpha}}
\label{L_max}
\end{equation}

\noindent where $\lambda_{max}$ is the most unstable wavelength. Given the adopted assumptions, the transverse wavenumber $k_{trans}$ and the transverse wavelength $\lambda_{trans}$ of the three-dimensional ripples scale, respectively, with Eqs. \ref{k_max} and \ref{L_max}.

Estimations of $\Omega_i$ and $\lambda_{max}$ depend on estimations of $\beta$. This is done next for the case of aquatic ripples, in order to compare $\lambda_{max}$ with experimental data presented in Section \ref{section:comparison}.

If we consider that $10^{-4}\leq \xi /R\leq 10^{-1}$, which is in accordance with $\xi\ll R$, then $\log\left( \frac{\xi}{R} \right)\,=\,O(1)$. The order of the term

\begin{equation}
\Gamma\,=\,\int{\vec{\omega}\cdot\vec{dA}}
\label{circulation}
\end{equation}

\noindent where $\vec{\omega}$ is the vorticity in the $s$ direction and $\vec{dA}$ is the element of the transverse area $A$, can be evaluated as follows. Let the size of the vortex scale with the height $H\approx 10^{-2}m$ of observed ripples and let the shear flow velocity be $U_S\approx 0.1m/s$. In this case,

\begin{equation}
\left|\vec{\omega}\right|\,\sim\,\frac{\partial U_S}{\partial y}\,\approx\,1s^{-1}
\label{vorticity}
\end{equation}

\noindent and $A\,\sim\,H^2\,\approx\,10^{-4}m^2$. The circulation is then $\Gamma\,\approx\,10^{-4}m^2/s$ and $\beta\,\approx\,10^{-5}m^2/s$. Finally, for normalizing purposes, consider that the mean grain diameter is $d\,\approx\,10^{-4}m$ and that $\omega^{-1}$ is a relevant time scale. 

\begin{figure}
  \begin{center}
    \includegraphics[width=0.6\columnwidth]{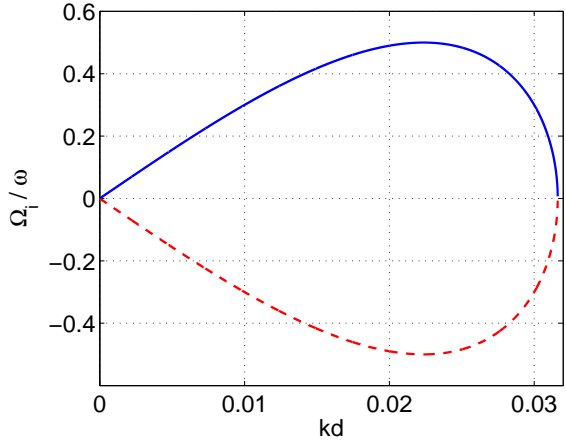}
    \caption{Dimensionless growth rate $\Omega_i /\omega$ as a function of the dimensionless wavenumber $kd$. The continuous line corresponds to $\Omega_i>0$ and the dashed line corresponds to $\Omega_i<0$.}
    \label{fig:growth}
  \end{center}
\end{figure}

Figure \ref{fig:growth} presents the dimensionless growth rate $\Omega_i /\omega$ as a function of the dimensionless wavenumber $kd$ for $\alpha\,=\,1s^{-1}$. This figure shows the stable $\Omega_i<0$ and unstable $\Omega_i>0$ branches of the solution, and it is valid only for the long-waves. As discussed in obtaining Eq. \ref{k_max}, the unstable branch has a maximum, which corresponds to the most unstable mode. For this mode, Eq. \ref{k_max} predicts $kd\,=\,0.0224$, in accordance with Fig. \ref{fig:growth}. Equation \ref{L_max} predicts $\lambda_{max}/d\,=\,281$.

\section{COMPARISON WITH EXPERIMENTS}
\label{section:comparison}

This section compares the present analysis with some experimental results, mainly that of \cite{Franklin_3}. In accordance with the assumptions of the model, the wavelength predicted for the line vortex is compared to the transverse length-scale of ripples and dunes. Given the large dispersion of the experimental data, the present comparison is limited to the orders of magnitude of the wavelengths.

\begin{figure}
  \begin{center}
    \includegraphics[width=0.5\columnwidth]{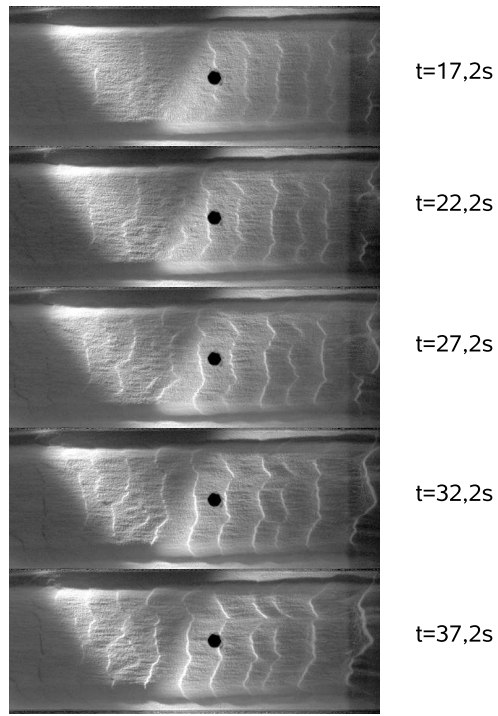}
    \caption{Evolution of ripples from an initially flat granular bed sheared by a turbulent water flow (top view). Flow direction is from right to left, $U=0.35m/s$ and the granular bed is composed of zirconium beads with $d=0.18mm$. Figure extracted from \cite{Franklin_3}.}
    \label{fig:inst_franklin}
  \end{center}
\end{figure}

Franklin (2008) \cite{Franklin_3} experimentally studied the initial instabilities on different granular beds under turbulent water flows. The experimental test section was a $6m$ long horizontal closed-conduit of rectangular cross-section ($120mm$ wide by $60mm$ high), made of transparent material. The employed fluid was water and the granular bed was composed of glass and zirconium beads. The cross-section averaged velocity of the water was in the range $0.2m/s\,<\,U\,<\,0.4m/s$ and the mean grain diameter $0.1mm\,<\,d\,<0.6mm$. Franklin (2008) \cite{Franklin_3} measurements showed that the initial bedforms are two-dimensional ripples. After the initial phase, bedforms evolve to three-dimensional ripples, as seen in Fig. \ref{fig:inst_franklin}.

Considering all the measurements presented in \cite{Franklin_3}, the longitudinal wavelength of initial ripples was in the range $90\,<\,\lambda_{long}/d\,<250$. Franklin (2008) \cite{Franklin_3} observed that the transverse length-scale $\lambda_{trans}$ of three-dimensional ripples was $O(\lambda_{long})$, so that $\lambda_{trans}/\lambda_{long}=O(1)$. These results are in accordance with the wavelength of the vortex instability predicted by Eq.\ref{L_max} in Section \ref{section:analysis}, where the order of $\beta$ was estimated based on the conditions reported in \cite{Franklin_3}.

The experimental observation that $\lambda_{trans}/\lambda_{long}=O(1)$, together with the fact that the two-dimensional ripples result from a long-wavelength instability, is an \textit{a posteriori} justification for the long-wavelength approach adopted in Section \ref{section:analysis}.

Franklin (2008) \cite{Franklin_3} also observed that, whenever the granular bed is not thick enough, the three-dimensional ripples evolve into isolated ripples and dunes that keep, at least initially, their longitudinal and transverse length-scales. If we consider this, the $\lambda_{trans}/\lambda_{long}=O(1)$ prediction can be compared to proposed relations between the length and the width of isolated dunes. In general, these relations propose that $\lambda_{long}$ and $\lambda_{trans}$ have the same order. For example, based on measurements of isolated dunes in Southern Morocco, Sauermann et al. (2000) \cite{Sauermann_1} proposed that $\lambda_{trans}\,\approx\,0.8\lambda_{long}$. In a review paper, Andreotti et al. (2002) \cite{Andreotti_1} proposed, based on experimental and field data, that $\lambda_{trans}\,\approx\,1.3\lambda_{long}$. Based on experiments, Franklin and Charru (2011) \cite{Franklin_7} proposed that $\lambda_{trans}\,\approx\,0.8\lambda_{long}$ for aquatic isolated dunes.

It is important to note here that the aquatic ripples are the equivalent of the aeolian dunes: they are both shaped by the direct perturbation of the fluid flow \cite{Andreotti_1, Andreotti_2, Claudin_Andreotti}. This justifies the comparisons made in the above paragraph. The nomenclature is misleading, the aeolian ripples being the result of granular impacts (and of reptation), and not directly related to fluid flow perturbations \cite{Andreotti_3}.

The cited works seem to corroborate that $\lambda_{trans}\,\sim\,\lambda_{max}$. In this case, the three-dimensional patterns are the product of the vortex instability, and the analysis presented in Section \ref{section:analysis} can be used to determine $\lambda_{trans}$.

\section{CONCLUSIONS}
\label{section:conclusions}

This paper presented a model to predict the transverse length-scale of three-dimensional ripples. It is based on the theoretical and experimental evidence that the initial bedforms are two-dimensional and that, during their growth, a recirculation vortex is formed downstream of their crests. The model then explores the idea proposed by Raudkivi (2006) \cite{Raudkivi_2} that deformations of the downstream vortex lead to three-dimensional ripples.

The recirculation vortex was modeled as a line vortex sheared by an external flow, and a stability analysis was conducted. The transverse length of three-dimensional ripples was supposed to scale with the most unstable wavelength predicted for the vortex. The obtained transverse scale compared well to experimental data of three-dimensional ripples and dunes.

\section{ACKNOWLEDGMENTS}

The author is grateful to Petrobras S.A. for the financial support to write this article (contract number 0050.0045763.08.4).






\bibliography{references}
\bibliographystyle{elsart-num}







\end{document}